\documentclass[conference]{IEEEtran}
\IEEEoverridecommandlockouts

\usepackage{cite}
\usepackage{amsmath,amssymb,amsfonts}
\usepackage{algorithmic}
\usepackage{graphicx}
\usepackage{textcomp}
\usepackage{xcolor}
\def\BibTeX{{\rm B\kern-.05em{\sc i\kern-.025em b}\kern-.08em
    T\kern-.1667em\lower.7ex\hbox{E}\kern-.125emX}}

\usepackage{algorithmic}
\usepackage{graphicx}
\usepackage{textcomp}
\usepackage{xcolor}
\usepackage{subcaption}
\usepackage{threeparttable}
\usepackage{url}

\usepackage{mdwtab}
\usepackage{booktabs}
\usepackage{array}
\usepackage{mdwmath}
\usepackage{eqparbox}
\usepackage{multirow, multicol}
\usepackage{makecell}
\usepackage{tabularx}
\usepackage{diagbox}
\usepackage[inkscapelatex=false]{svg}
\usepackage{siunitx}
\usepackage{verbatim}
\usepackage{pifont} 
\usepackage{bbding}
\usepackage{svg}

\newcommand\like[1]{\begin{picture}(0.6,0.6)
\ifnum0=#1\put(0.3,0.25){\circle{0.6}}\else
\ifnum10=#1\put(0.3,0.25){\circle*{0.6}}\else
\put(0.3,0.25){\circle{0.6}}\put(0.3,0.25){\circle*{.3}}
\fi\fi\end{picture}}

\usepackage[letterpaper, bottom=1.1in, top=0.755in, left=0.755in, right=0.755in]{geometry}

\begin{document}


\title{SoK: Credential-Based Trust Management in Decentralized Ledger Systems}

\author{Yanna Jiang$^1$, Haiyu Deng$^1\textsuperscript{\,\Envelope}$, Qin Wang$^2$, Guangsheng Yu$^1$, Xu Wang$^1$, Yilin Sai$^2$,\\ Shiping Chen$^2$, Wei Ni$^1$, Ren Ping Liu$^1$
\\
\textit{$^1$University of Technology Sydney} $|$
\textit{$^2$CSIRO Data61}
}


\maketitle

\begin{abstract}
Trust management systems (TMS) are crucial for managing trust in distributed environments. The rise of decentralized systems and blockchain has sparked interest in credential-based decentralized trust management systems (DTMS). This paper bridges the gap between theory and practice through a systematic review of credential-based DTMS. We analyze existing DTMS solutions through multiple dimensions, including their architectural designs, credential mechanisms, and trust evaluation models. Our survey provides a detailed taxonomy of credential-based DTMS approaches and establishes comprehensive evaluation criteria for assessing DTMS implementations. Through extensive analysis of current systems and implementations, we identify critical challenges and promising research directions in the field. Our examination offers valuable insights for researchers and practitioners working on DTMS,  particularly in areas such as access control, reputation systems, and blockchain-based trust frameworks.
\end{abstract}


\section{Introduction} \label{introduction}

With the rise of Internet, a growing number of individuals and organisations are connected with each other to share information and services. The emergence of Internet of Things (IoT) and blockchain accelerate the trend, enabling networks that are more diverse, dynamic, and open.

The nature of such networks brings new security challenges, as interacting entities may be unfamiliar with each other, rendering identity-based security mechanisms inadequate for regulating interactions \cite{Ardagna2007}.
Trust Management Systems (TMS) \cite{Blaze1996} were introduced to enhance network security by enabling entities to assess whether another entity is qualified to provide or utilize specific information or services \cite{Ardagna2007, Franceschi2021}. This assessment is based on evaluating the trustworthiness of entities using various evidence and criteria \cite{Opiola2018, Wei2021}. 
TMS have garnered significant interest from the research community and have been applied in areas such as access control and authorization \cite{Blaze1996, Yan2017, Andersen2019, Lauinger2021}, identity and data authentication \cite{Zhao2022, More2021, Yu2022, Javaid2020}, malicious node detection \cite{Boudagdigue2020, Baudet2022, Nichols2021}, and the prevention of internal attacks \cite{Wang2022, Baudet2022, Li2022}, including attacks by authenticated but hostile entities, such as Sybil attacks. 
In addition, TMS has been applied to enhance service quality of networks by optimizing routing path selection \cite{Chen2014, AIREHROUR2019860}, service provider selection \cite{Wei2021, Chang2005, Awan2019}, and resource allocation \cite{Wang2020, Cui2021}.

TMS can be classified into credential-based and behavior-based systems \cite{Cho2011}, with the former verifying entity credentials \cite{Opiola2018} and the latter assessing historical behaviors \cite{Wei2021}. While credential-based TMS provide robust security analysis but face dynamicity challenges, behavior-based systems offer dynamic evaluation but suffer from cold start problems and security uncertainties \cite{Bampatsikos2021, Opiola2018}. This paper focuses more on credential-based TMS.

Architecturally, TMS can be implemented in centralised, decentralised, or hybrid architectures \cite{FanXinxin2021}. Centralized systems rely on a central authority for trust management through credential stores \cite{ThummalaC15} or periodic evaluations \cite{Chen2019}, but face issues with single points of failure and scalability \cite{WangJie2022, Andersen2019}. In contrast, decentralized TMS (DTMS) enable autonomous trust management by individual entities, improving scalability and robustness while aligning with trust's inherent subjectivity \cite{Blaze1996}. With the emergence of Blockchain \cite{cao2022blockchain,liu2024distributed} and Web3 \cite{wang2022exploring, liu2023web3}, DTMS have gained significant traction \cite{Andersen2019, Franceschi2021} and form the focus of this paper.

Despite growing interest in DTMS, existing research remains fragmented across three dimensions: (a) Studies predominantly focus on specific technical implementations (e.g., blockchain-based solutions \cite{Andersen2019}) without systematic analysis of architectural patterns; (b) Current classifications lack multi-layered perspectives integrating governance models with technical implementations \cite{Franceschi2021}; (c) Evaluation criteria for dynamicity and explainability remain inconsistent across domains \cite{WangJie2022}. 
We address these gaps with the following contributions.
\begin{itemize}
    \item We provide a detailed analysis of trust characteristics, including dynamicity and explainability, and their implications for system design, forming the theoretical foundation for credential-based DTMS.
    
    \item We conduct a comprehensive classification and layered analysis of existing credential-based DTMS, focusing on trust formation, trust verification, and trust management, to highlight the integration of governance, network, and data layers for enhanced system interoperability.
    
    \item We establish a set of evaluation criteria focusing on dynamicity, privacy, scalability, and explainability, offering a systematic approach to assess the effectiveness of credential-based DTMS implementations.
    
\end{itemize}
    
The rest of this paper is organized as follows. Section~\ref{description} explores trust characteristics and system design implications, Section~\ref{survey} analyzes credential-based DTMS, Section~\ref{evaluation} presents evaluation criteria, and Sections~\ref{challenge} and~\ref{conclusion} discuss future directions and conclusions.

\section{Trust System Foundations}\label{description}

\subsection{Trust} \label{trust}
Trust originated from social sciences has been adapted to computer science and network security \cite{Li2022}. In distributed systems, trust represents a quantifiable belief about an entity's capability to perform specific tasks \cite{Li2007, Sagar2020}.

Trust has several key characteristics that can be divided into two categories. The first category includes implicitness, transitivity, asymmetry, antonymy, asynchrony, and gravity \cite{Chang2005, Ebrahimi2022, Pranata2012AHR}. The second includes subjectivity, dynamicity, and context-awareness \cite{WangJie2022, Ebrahimi2022}. These characteristics are interconnected - for example, subjectivity influences asymmetry and gravity, while dynamicity emerges from transitivity and context-awareness.

\smallskip
\subsubsection{Implicitness} \label{implicitness}
Trust exhibits implicitness in two ways. First, entities cannot easily quantify trust due to complex evidence \cite{Chang2005, WangJie2022}. Second, trust emerges implicitly through actions \cite{Opiola2018}, such as granting access permissions. Security systems frequently rely on the implicit trust assumptions \cite{Jensen2014}.

\subsubsection{Transitivity}
Trust can propagate: when A trusts B, A may also trust entities trusted by B. This indirect trust \cite{Ebrahimi2022} enables delegation \cite{Andersen2019} and underlies graph-based trust models like SDSI/SPKI \cite{Rivest1996, Ellison1999}. It is particularly important in service-oriented architectures for provider selection \cite{Chang2005}.

\subsubsection{Dynamicity} \label{dynamicity}
Trust relationships evolve over time based on context \cite{Yosra2013, Ebrahimi2022, Rafey2016}, including entity states and environmental conditions \cite{Lin2018}. Trust levels can increase through positive interactions or decay naturally \cite{Cho2015, WangJie2022}. Modern systems implement this through mechanisms like PKI certificate revocation \cite{Myers1999} and IoT trust updates \cite{Huang2021, Chen2016, Rafey2016}.

\subsubsection{Subjectivity} \label{Subjectivity}
Trust is inherently subjective \cite{Audun2007}, with each entity having its own standards for establishing and evaluating trust. This manifests through different access to evidence \cite{Cho2015} and entity-specific evaluation criteria \cite{Ahmed2019}. This subjectivity is fundamental to DTMS design \cite{Blaze1996}, enabling local control over trust relationships \cite{WangJie2022, Opiola2018, Andersen2019}.

\smallskip
These characteristics shape the design of trust-based systems, presenting both challenges and opportunities for implementing novel trust mechanisms.

\subsection{Trust, Reputation, and Risk}

In behavior-based DTMS, trust and reputation are related but distinct. Trust is a subjective, entity-driven assessment, while reputation reflects a passive, collective evaluation from the community~\cite{Chen2011TRMIoTAT, JOSANG2007618, WangJie2022}. This distinction allows trust to diverge from reputation in specific scenarios.


Some DTMS architectures incorporate reputation-based trust evaluation mechanisms \cite{AYED2023103093}, potentially leading to centralization through dominant highly-trusted entities \cite{Wang2022}. This centralization stems from a positive feedback loop where elevated reputation scores reinforce trust levels, which enhances reputation. 
To maintain decentralization, implementations typically incorporate multiple trust evaluation factors, including direct interaction history.

Risk is a core element of trust. \cite{Avizienis2004} defines trust as an entity's willingness to depend on another without failure mitigation, framing trust as a decision to accept potential failure. \cite{Baudet2022} builds on this by linking trust levels to risk-benefit analysis, as shown in a decentralized PKI model. Risk assessment both informs trust decisions and clarifies underlying assumptions. In networks, it helps evaluate vulnerabilities. \cite{Lu2022} proposed a framework to assess attack impacts in trust-aware systems, showing how trust propagation \cite{Fan2016, Kamvar2003} can increase risk through the spread of trust scores.


\subsection{Trust Policy, Proof and Verifiability} \label{trust policy}


Trust policies are fundamental to TMS. Initially introduced for access control in early DTMS work~\cite{Ardagna2007}, they were formalized as predicates evaluating actions~\cite{Blaze1996} and implemented as unsigned credentials to define authority. As DTMS evolved, trust policies grew to include rules, algorithms, and parameters for credential-based trust decisions~\cite{More2021}. Recent approaches favor declarative expressions~\cite{Li2003, Nichols2021, Podder2022} to support security analysis and coordination.


Credential-based TMS employ proofs to demonstrate entity qualifications, subject to trust policy verification \cite{Lauinger2021, Franceschi2021, Salve2022}. These proofs manifest as credentials, cryptographic computations, or combinations thereof. Beyond enhancing verification efficiency, proof mechanisms enable privacy preservation via cryptographic techniques like zero-knowledge proofs.

The integration of trust policies and proof lifecycle establishes system verifiability. In credential-based DTMS, verifiability encompasses multiple dimensions:

\begin{itemize}
    \item \textbf{Mechanism-based}: Cryptographical (integrity verification), structural (semantic verification), and institutional (legitimacy verification) \cite{Nichols2021, Jeyakumar2022};
    \item \textbf{Object-based}: Hierarchical verification spanning identity \cite{Javaid2020}, cross-domain credentials \cite{Zhao2022}, role statements \cite{Salve2022}, and data packets \cite{Nichols2021}.
\end{itemize}

This hierarchical verification structure creates dependencies from higher-level (roles, data) to lower-level (credentials, identities) verifications \cite{Davie2019}.

Credential verification in DTMS operates peer-to-peer between prover and verifier \cite{Andersen2019}, without issuer involvement. This decentralized model, supported by cryptography and DLT, enables autonomous trust management and verification~\cite{Jeyakumar2022}.

\section{Credential-Based DTMS Framework} \label{survey}


DTMS require a systematic framework to analyze their architectures and functionalities. 
While existing research has proposed various approaches to decompose trust systems - such as the six-phase trust computation model for IoT \cite{Guo2017} and the four-layer Decentralized Identifier (DID)-based trust stack \cite{Davie2019} - these frameworks are either too specific to certain domains or lack consideration of the distributed nature of modern trust systems. To address this gap, we propose a two-dimensional framework specifically designed for analyzing DTMS, where one dimension is sequential and the other is spatial. The sequential dimension captures trust lifecycle phases while the spatial dimension describes architectural layers. This comprehensive framework enables systematic classification and analysis of state-of-the-art DTMS solutions.

\begin{figure}[h!]
     \centering
     \begin{subfigure}[b]{0.3\linewidth}
         \centering
         \includegraphics[width=\textwidth]{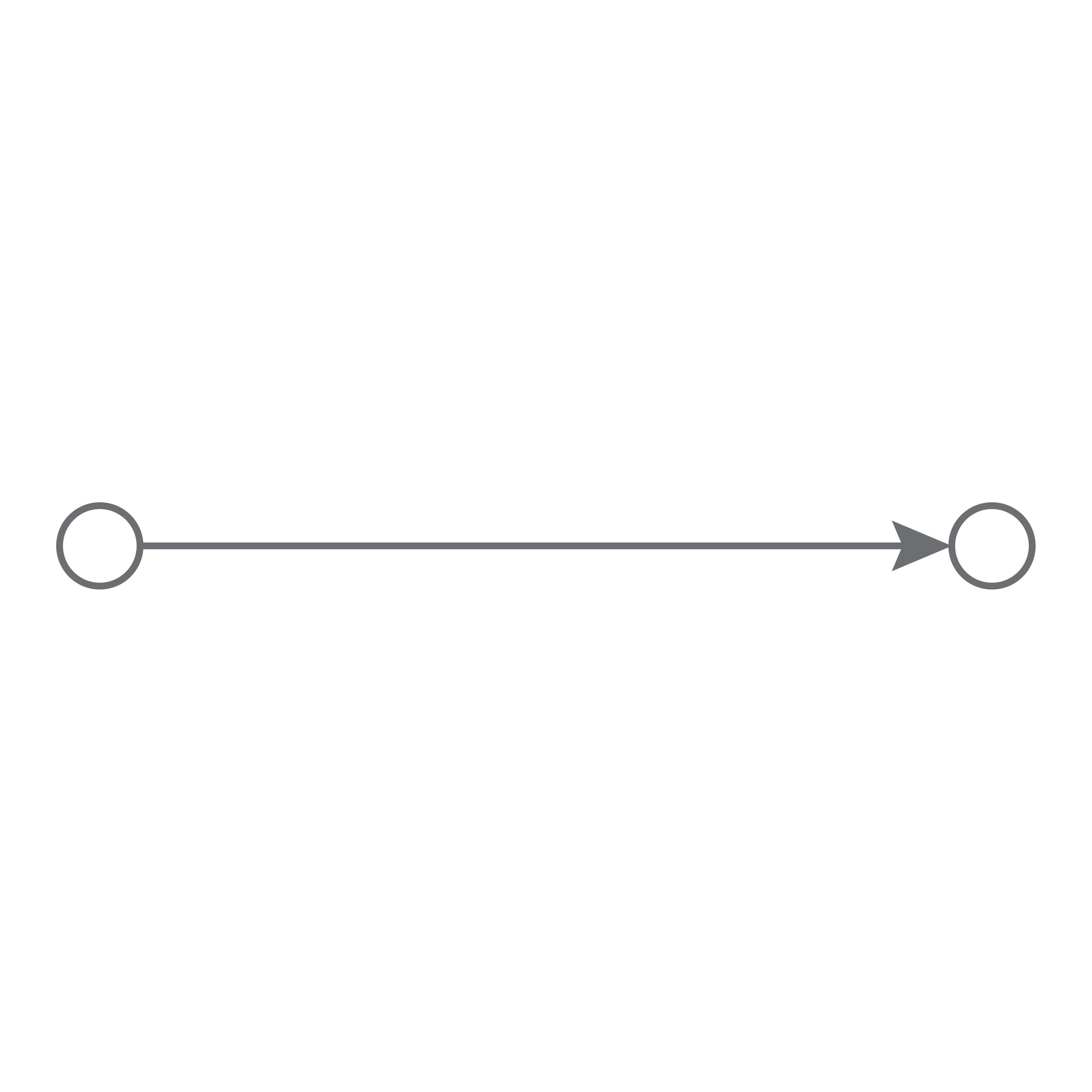}
         \caption{\footnotesize{Trust Formation}}
         \label{fig:Trust Formation}
     \end{subfigure}
     \hfill
     \begin{subfigure}[b]{0.3\linewidth}
         \centering
         \includegraphics[width=\textwidth]{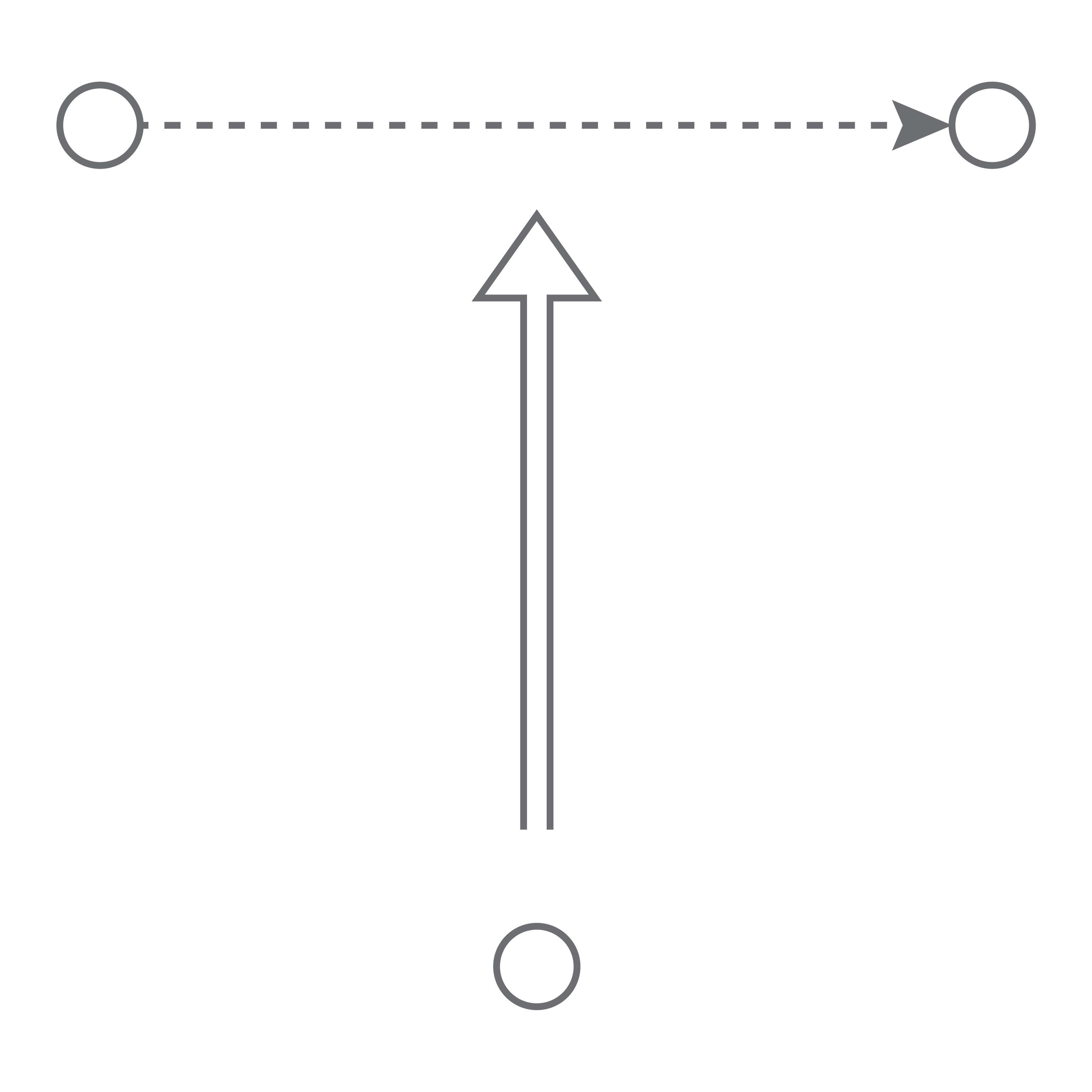}
         \caption{\footnotesize{Trust Verification}}
         \label{fig:Trust Verification}
     \end{subfigure}
     \hfill
     \begin{subfigure}[b]{0.3\linewidth}
         \centering
         \includegraphics[width=\textwidth]{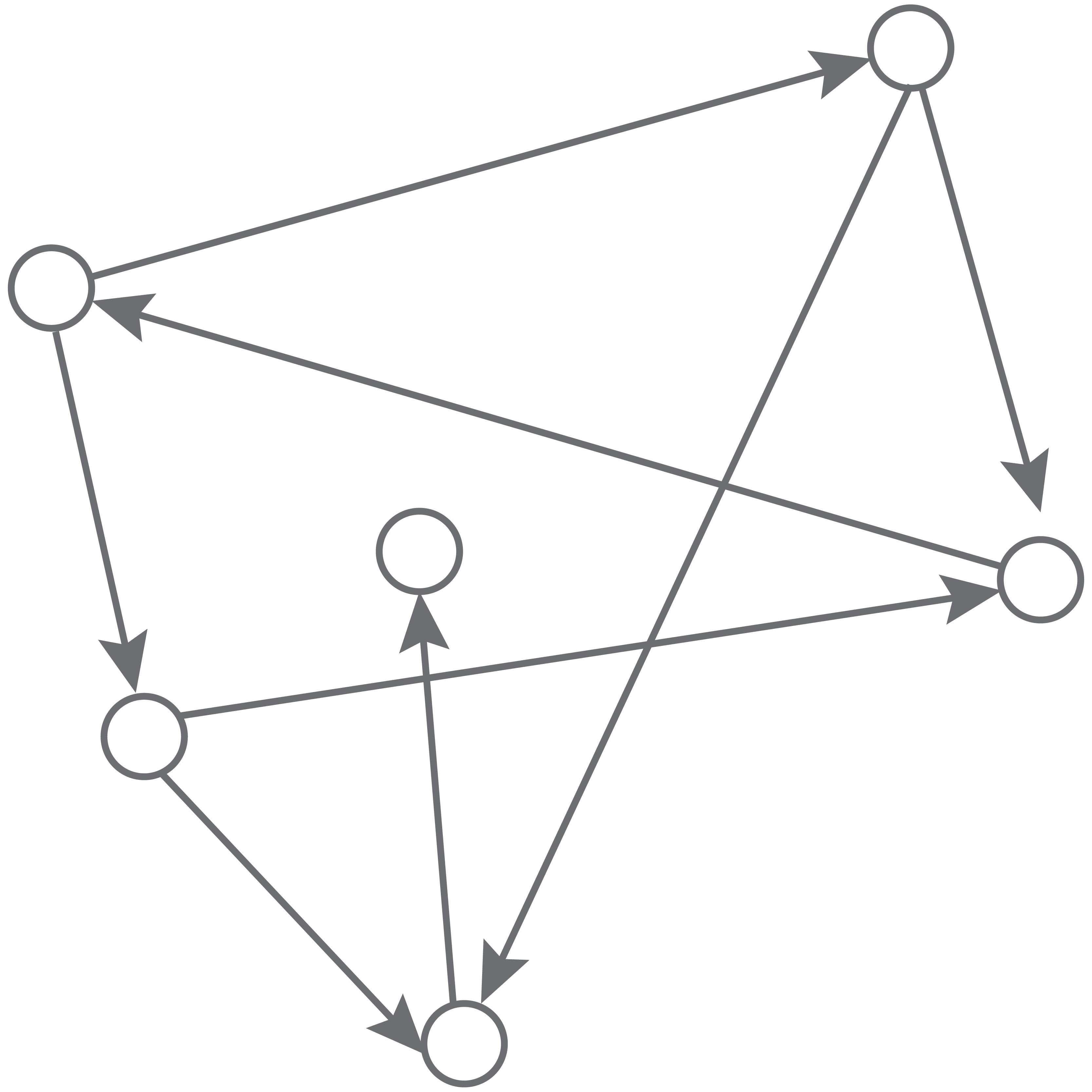}
         \caption{\footnotesize{Trust Management}}
         \label{fig:Trust Management}
     \end{subfigure}
        \caption{Three-Phase Sequential Classification}
        \label{fig:tasks}
\end{figure}

For the sequential dimension, as illustrated in Figure \ref{fig:tasks}, DTMS encompasses three fundamental phases:
\begin{itemize}
    \item \textbf{Trust formation}. Trust formation focuses on establishing and evaluating trust relationships through empirical evidence, including behavioral history, consensus mechanisms, and existing trust networks. This phase facilitates trust establishment between a trustor and trustee through systematic evaluation methodologies.
    
    \item \textbf{Trust verification}. Trust verification enables cryptographic validation of established trust relationships by third parties. This phase implements verification protocols that allow external verifiers to authenticate the existence and validity of trust relationships.
    
    \item \textbf{Trust management}. Trust management encompasses the systematic maintenance and governance of trust relationships, including dynamic updates, certificate lifecycle management, and cross-domain interoperability. This phase ensures network sustainability through robust trust infrastructure maintenance.
\end{itemize}

\begin{figure}[h!]
\centering
\includegraphics[scale=0.2]{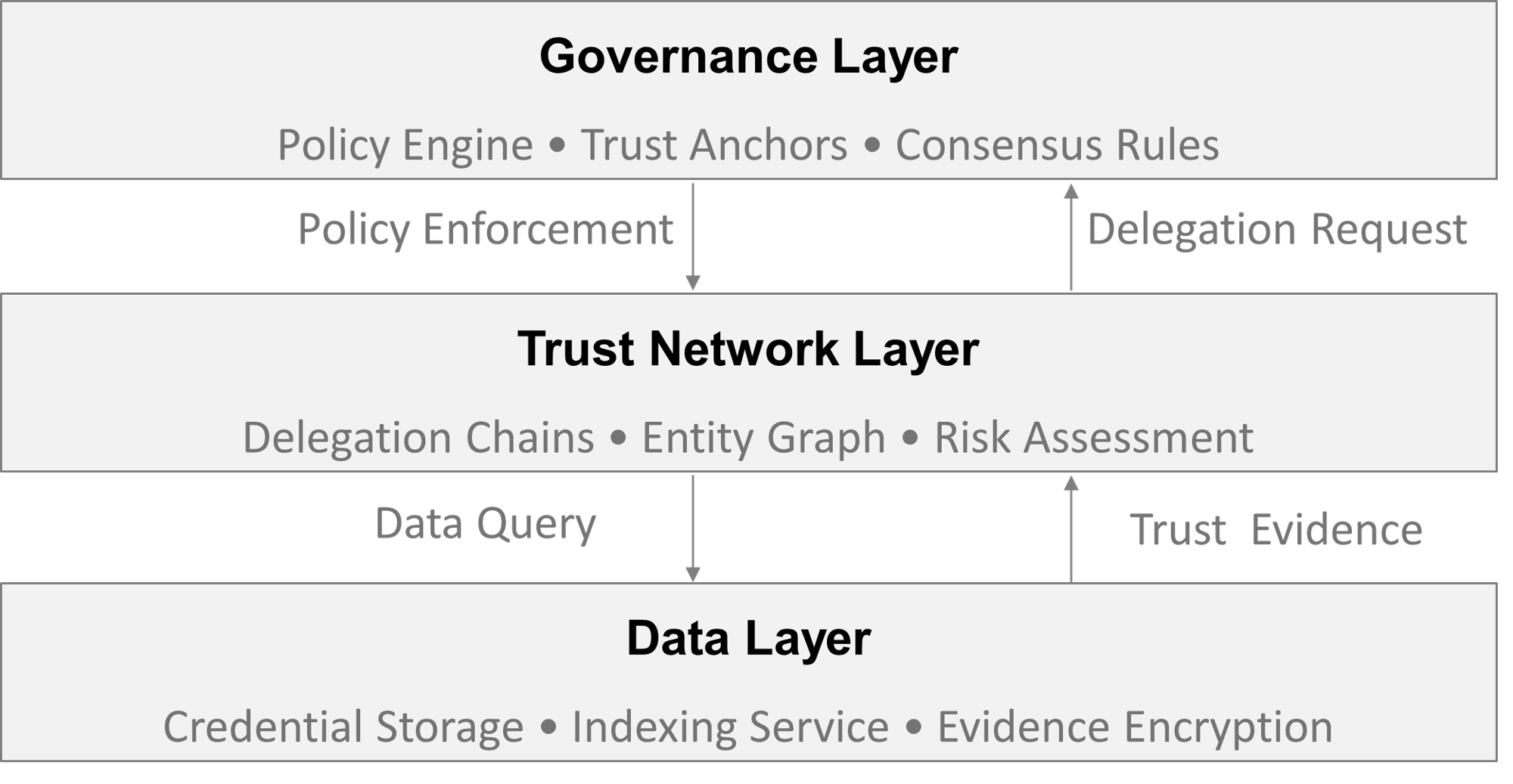}
\caption{Three-Layer Spatial Classification}
\label{fig:layers}
\end{figure}

For the spatial dimension, as illustrated in Figure \ref{fig:layers}, DTMS can be decomposed into three distinct architectural layers:
\begin{itemize}
    \item \textbf{Data layer.} The data layer focuses on the fundamental infrastructure for credential management, including efficient mechanisms for indexing, querying, and distributing trust-related data~\cite{nguyen2025blockchain}. This layer ensures data integrity, privacy preservation, and non-repudiation through cryptographic primitives and secure storage protocols.
    
    \item \textbf{Trust network layer.} The trust network layer addresses the topology and structural aspects of trust relationships. By modeling entities and their interactions as a network, this layer enables sophisticated trust evaluation and verification through graph-theoretic analysis \cite{li2020direct,wang2023sok} and network algorithms.
    
    \item \textbf{Governance layer.} The governance layer establishes protocol specifications and consensus mechanisms that regulate network operations. This includes defining trust anchor roles, authority delegation protocols, and election procedures for administrators. The layer provides a standardized framework for decentralized trust management.
\end{itemize}

Table~\ref{paper_table} presents a systematic three-by-three classification matrix that integrates temporal phases and architectural layers, enabling comprehensive analysis of DTMS solutions.


\begin{table*}[t]
\renewcommand{\arraystretch}{1.1}
\caption{Two-dimensional Classification of Credential-based DTMS Papers}
\label{paper_table}
\centering
\begin{tabularx}{\textwidth}{|>{\centering\arraybackslash}X 
  | >{\centering\arraybackslash}X
  | >{\centering\arraybackslash}X
  | >{\centering\arraybackslash}X|}
\hline
Layer  & \textbf{Trust Formation} & \textbf{Trust Verification} &  \textbf{Trust Management} \\
\hline
\hline
\textbf{Governance}  & \cite{Sarker2021}, \cite{Baudet2022} & \cite{Zhao2022}, \cite{Nichols2021} & \cite{Yu2022}, \cite{Boudagdigue2020} , \cite{10648633},\cite{xiang2024blockchain}\\
\cline{2-4}
\textbf{Trust Network} & \cite{Wang2022} & \cite{Franceschi2021}, \cite{Salve2022}, \cite{10819652} & \cite{10528325}\\
\cline{2-4}
\textbf{Data} & \cite{Wang2022}, \cite{Baudet2022} & \cite{Lauinger2021}, \cite{Franceschi2021}, \cite{Zhao2022}, \cite{Nichols2021}, \cite{Javaid2020} & \cite{More2021}, \cite{Yu2022} \\
\hline
\end{tabularx}
\end{table*}


\smallskip
\subsection{Trust Formation}
Trust formation is a critical phase in DTMS that focuses on establishing and evaluating initial trust relationships between entities. This section examines three key approaches: Certificate Authority (CA)-based trust evaluation in maritime networks, blockchain-based PKI root management, and decentralized multi-agent PKI. These solutions demonstrate how trust can be established through hierarchical CA structures, democratic blockchain voting, and distributed agent-based systems respectively.

\smallskip
\subsubsection{CA-Based Trust Evaluation in Maritime Networks}

Wang \textit{et al.}~\cite{Wang2022} proposed a hierarchical CA-based architecture for trust evaluation in Maritime Autonomous Surface Ship (MASS) networks. Root CAs delegate to intermediate CAs, forming CA trees, and each entity receives a certificate for secure communication.

At the trust network layer, trust is computed as \(T(entity_{i} \rightarrow entity_{j})=\lambda^{DTL(entity_{j}|entity_{i})}\), where \(\lambda\in(0,1)\) and DTL is the shortest path to the nearest common CA. Initial trust is based on CA topology and dynamically updated through interaction behaviors.

Entities manage Inherent Trains of Trusted CAs (ITTCs), which are exchanged via gossip protocol to compute trust degrees. Results are stored on a blockchain through a trust-aware consensus. Simulations show effectiveness in isolating malicious entities and supporting cross-tree trust, though certificate updates, privacy, and scalability require further investigation.




\smallskip
\subsubsection{Blockchain-Based PKI Root Management}

Sarker \textit{et al.}~\cite{Sarker2021} proposed a Blockchain-Based Root Management (BBRM) scheme to enhance the US DOT's Security Credential Management System by improving transparency and resilience over the Elector-Based Root Management (EBRM) model. The governance layer enables democratic root CA control via smart contracts, where electors vote on Root Certificate (RC) operations and manage membership through Elector Certificates (ECs).

Implemented on Hyperledger Fabric, the system showed feasible performance with up to 9 electors. BBRM supports dynamicity and explainability through blockchain voting records, though privacy remains an open issue.



\smallskip
\subsubsection{Decentralized Multi-Agent PKI}
Baudet \textit{et al.} \cite{Baudet2022} introduced MAKI (Multi-Agent Key Infrastructure), a decentralized PKI for open multi-agent systems. MAKI eliminates the need for external CAs by distributing CA responsibilities among network agents.



The governance framework allows agent self-nomination as CAs based on local CA density, with trust levels (Low/Moderate/High) derived from direct and indirect signals. MAKI extends X.509 certificates to support decentralization, and its Sybil resistance is achieved through trust limitations on newcomers~\cite{qin2020cecoin}. While dynamic trust evaluation is supported, the absence of persistent trust evidence limits decision verifiability, and privacy and scalability remain open issues.

\smallskip
\subsubsection{Summary}
Trust formation solutions in credential-based DTMS demonstrate strong architectural foundations but share common challenges. Privacy concerns, especially around sensitive trust metadata, require further analysis. Scalability and performance evaluations are also often limited, particularly for solutions involving complex trust computations or blockchain components.


\subsection{Trust Verification}
Trust verification is a critical phase in DTMS that focuses on cryptographically validating established trust relationships. This section examines six key approaches, showing how trust can be verified through blockchain consensus, trust schemas, graph-based algorithms, cryptographic accumulators, decentralized identifiers, and hardware security modules.

\smallskip



\subsubsection{Blockchain-Based Cross-Domain Authentication}

Zhao \textit{et al.}~\cite{Zhao2022} proposed a double-layer blockchain architecture to enhance cross-domain authentication efficiency over traditional PKI systems. At the governance layer, Authentication Servers (AS) manage a consortium blockchain for verifying blockchain-based certificates, whose hashes serve as cross-domain proofs.

The data layer simplifies X.509 certificates by removing issuer and signature fields, relying on hash verification for integrity. A private blockchain manages intra-domain certificates, while the consortium chain handles cross-domain verification. Implementation on Hyperledger Fabric showed improved efficiency and scalability, though the explainability of verification decisions remains a challenge.




\smallskip
\subsubsection{Schema-Based IoT Access Control}

Nichols~\cite{Nichols2021} proposed an Information-Centric Networking architecture for IoT that achieves fine-grained access control using trust schemas. Built on Named Data Networking, the system secures data directly rather than communication channels.

At the data layer, packets follow a structured naming format including capability, topic, location, and issuer. The governance layer uses Versec DSL to define verification rules for different devices. A prototype showed feasible performance across CPUs. While the architecture supports dynamic policy updates and encrypted privacy, it lacks explainability due to non-persistent trust schemas.

\smallskip
\subsubsection{Role-Based Trust Verification}
Franceschi \textit{et al.} \cite{Franceschi2021} proposed DART (Decentralized Authority Recognition with Trust), a blockchain implementation of the role-based Trust management framework. DART manages RT$^0$ credentials via smart contracts and implements weighted trust evaluation.

At the trust network layer, DART employs a backward search algorithm to build proof graphs for role verification, utilizing trust weights to calculate cumulative trust across credential chains. However, executing this algorithm on-chain has shown significant computational overhead on Ethereum.

To address DART's scalability issues, De Salve \textit{et al.} \cite{Salve2022} proposed L2DART as a layer-2 solution. L2DART moves graph computation off-chain while maintaining on-chain verification of proof paths. This hybrid approach significantly reduced gas costs while preserving auditability.

\smallskip
\subsubsection{Anonymous Credential Authorization}
Lauinger \textit{et al.} \cite{Lauinger2021} addressed credential issuer verification in Self-Sovereign Identity Management (SSIM) systems through Anonymous Proof of Authorization (A-PoA). The scheme enables Root Authorities (RA) to authorize Certificate Issuing Authorities (CIA) while preserving CIA anonymity.

At the data layer, the solution employs Rivest-Shamir-Adleman (RSA) accumulators for membership verification and Non-interactive Zero Knowledge Proof of Knowledge of Exponent (NI-ZKPoKE) for anonymous proofs. The protocol achieves O(1) verification complexity while maintaining O(n) complexity for authorization operations.

\smallskip
\subsubsection{DID Trust} 
Yin \textit{et al.} \cite{10819652} proposed DidTrust, a novel decentralized identity trust verification protocol that focuses on robust trust verification while achieving privacy preservation and attack resilience. 
At the trust network layer, DidTrust models trust relationships through a directed graph structure where vertices represent DID entities and edges capture trust attestations with associated feedback data. 
Trust values are aggregated using weighted path multiplication with configurable decay parameters to control propagation scope. Experimental results demonstrated superior attack detection rates compared to existing approaches while maintaining strong privacy guarantees through UC-secure protocols.

\smallskip
\subsubsection{Hardware-Based Vehicle Authentication}
Javaid \textit{et al.} \cite{Javaid2020} proposed a blockchain-based authentication protocol for Internet of Vehicles (IoV) networks using Physical Unclonable Functions (PUF). The protocol leverages PUF-derived challenge-response pairs as hardware fingerprints for vehicle authentication.

At the data layer, vehicle identities combine blockchain addresses with PUF credentials~\cite{10.1145/3656166}. Road Side Units (RSU) verify vehicles through challenge-response protocols and maintain certificates on-chain. Implementation with dynamic Proof of Work (dPoW) demonstrated 36\% lower communication overhead compared to existing approaches.

\smallskip
\subsubsection{Summary}
Trust verification solutions demonstrate strong scalability support, particularly in blockchain-based approaches. Solutions at the trust network layer face dynamicity challenges, while simpler credential models generally achieve better capability coverage than graph-based approaches. Future work should address privacy preservation in trust networks and explainability of verification decisions.

\subsection{Trust Management}
Trust management focuses on maintaining and governing established trust relationships through systematic approaches. This section examines four key solutions: blockchain-based trust certification, schema-based certificate revocation, game-theoretic certificate management, and Sovereign Identity (SSI) integration. These approaches demonstrate how trust can be managed through decentralized trust networks, structured naming conventions, and behavior-based evaluation respectively.

\smallskip
\subsubsection{Blockchain-Based Trust Certification}
Chen et al. \cite{10648633} propose a consortium blockchain-based DTMS featuring cross-layer trust evaluation between vehicles and RSUs, TEE-accelerated VDF consensus for efficient synchronization, and attack-resistant smart contracts for behavioral pattern analysis. More \textit{et al.} \cite{More2021} proposed a blockchain-based Web of Trust (WoT) system that enables decentralized certification of credential issuer legitimacy. The system implements a graph-based trust model where trust certifiers publish verifiable statements about issuers.

At the data layer, trust statements are structured as tuples containing: certifier identity, issuer identity, legitimacy level \(LL\in[-1,1]\), confidence level \(LC\in[-1,1]\), digital signature, and timestamp. The system maintains two key graph structures: (1) A WoT graph where vertices represent entities and edges represent trust statements; (2) A transformation graph where vertices represent credential schema URIs and edges represent schema transformation rules.



The governance framework allows dynamic trust updates and revocation via timestamp-based supersession, with certifier authority validated through the Web of Trust. Ethereum implementation demonstrated sub-second performance on 10,000-edge graphs. Recent studies~\cite{xiang2024blockchain,10528325} extend this approach to cross-domain trust, though privacy remains an open issue.

\smallskip
\subsubsection{Schema-Based Certificate Revocation}
Yu \textit{et al.} \cite{Yu2022} introduced CertRevoke, a certificate revocation framework for Named Data Networking (NDN) that leverages structured naming conventions and trust schemas. The architecture organizes entities into trust zones, each managed by a trust anchor that issues certificates to zone members.

At the data layer, the framework defines:
\begin{itemize}
\item Structured naming conventions for certificates and revocation records;
\item Mapping rules between certificate names and corresponding revocation records;
\item Support for both issuer-initiated and owner-initiated revocation.
\end{itemize}

The governance layer enforces trust schemas that define validation rules for revocation records, while an in-network caching mechanism enhances performance by minimizing ledger interactions.

Experimental results demonstrated 45-84.5\% improvement in record-checking latency with caching. The framework provides strong support for scalability and privacy while maintaining explainability through blockchain records.

\smallskip
\subsubsection{Game-Theoretic Certificate Management}

Boudagdigue \textit{et al.}~\cite{Boudagdigue2020} proposed a cluster-based certificate management protocol for Industrial Internet of Things (IIoT) networks using game theory. Cluster Leaders (CLs) oversee member behavior and certificate lifecycles. At the governance layer, certificate revocation is modeled as a repeated signaling game between a CL with actions \{revoke, update\} and potentially Malicious Node (MN) with actions \{normal, malicious\}.

The framework calculates mixed-strategy Perfect Bayesian Equilibrium (PBE) using:
\begin{itemize}
\item \(\omega_{j}^{*}\): Equilibrium probability of malicious behavior;
\item \(\lambda_{j}^{*}\): Equilibrium probability of certificate revocation.
\end{itemize}

MATLAB simulations demonstrated accurate detection of malicious behavior across multiple scenarios. While the solution provides strong dynamicity via behavior-based evaluation, explainability is limited by a lack of parameter persistence.

\smallskip
\subsubsection{SSI Integration}






SSI represents the latest evolution in identity management, transitioning from centralized to user-centric control \cite{allen_2016}. It empowers individuals with full authority over their digital identities and credentials \cite{Jeyakumar2022}, offering enhanced privacy, scalability, and interoperability \cite{wang2023exploring}.

Credential and identity management lie at the core of DTMS. Modern architectures \cite{Zhao2022,Nichols2021} employ entity identification and credential-based trust metrics \cite{Wang2022}, as exemplified by the RT Framework \cite{Li2002}. SSI reinforces DTMS by offering a cryptographically sound and privacy-preserving identity infrastructure.

Conversely, DTMS addresses SSI’s limitations. While SSI ensures \textit{technical trust} through cryptographic proofs, it lacks \textit{institutional trust} for authority validation \cite{Jeyakumar2022}. DTMS frameworks such as \cite{Lauinger2021} complement SSI with governance and authorization mechanisms. Together, SSI and DTMS form a unified foundation for decentralized identity and trust, enhancing overall system trustworthiness.

\smallskip
\subsubsection{Summary}
Trust management solutions demonstrate robust explainability support, particularly in blockchain-based approaches. However, privacy preservation during data collection and comprehensive scalability analysis remain key challenges. Future work should address these gaps while maintaining the strong dynamicity and explainability characteristics of current solutions.

\begin{table*}[t]
  \caption{Evaluation of Credential-based DTMS Papers}
  \label{evaluation_table}
  \centering
  \renewcommand{\arraystretch}{1.1} 
  \begin{tabularx}{\textwidth}{ |>{\centering\arraybackslash}X 
    | >{\centering\arraybackslash}X
    | >{\centering\arraybackslash}X
    | >{\centering\arraybackslash}X
    | >{\centering\arraybackslash}X
    | >{\centering\arraybackslash}X
    | >{\centering\arraybackslash}X |}
  \hline
  \textbf{Task Category}  & \textbf{Reference} & \textbf{Layer Category} & \textbf{Dynamicity} & \textbf{Explainability} & \textbf{Privacy} & \textbf{Scalability} \\
  
  \hline
  \hline
  
  \multirow{3}{*}{\makecell{Trust\\Formation}} & \cite{Wang2022} & TrustNet,Data & \like{0} & \like{5} & - & - \\
  \cline{2-7}
   & \cite{Sarker2021} & Govern & \like{10} & \like{10} & - & \like{10} \\
  \cline{2-7}
   & \cite{Baudet2022} & Govern,Data & \like{10} & \like{0} & - & - \\
   
  \hline
  \hline  
  
  \multirow{7}{*}{\makecell{Trust\\Verification}} & \cite{Zhao2022} & Govern,Data & \like{5} & - & \like{10} & \like{10} \\
  \cline{2-7}
  & \cite{Nichols2021} & Govern,Data & \like{10} & \like{0} & \like{10} & \like{10} \\
  \cline{2-7}
  & \cite{Franceschi2021} & TrustNet,Data & \like{0} & \like{10} & \like{0} & \like{0} \\
  \cline{2-7}
  & \cite{Salve2022} & TrustNet & \like{0} & \like{10} & \like{0} & \like{10} \\
  \cline{2-7}
  & \cite{Lauinger2021} & Data & \like{10} & \like{5} & \like{10} & \like{5} \\
  \cline{2-7}
  & \cite{Javaid2020} & Data & \like{10} & \like{10} & \like{5} & \like{10} \\
  \cline{2-7}
  & \cite{10819652} & TrustNet,Data & \like{5} & \like{5} & \like{10} & \like{5} \\
  
  \hline
  \hline
  
  \multirow{5}{*}{\makecell{Trust\\Management}} & \cite{More2021} & Data & \like{10} & \like{10} & \like{0} & - \\
  \cline{2-7}
  & \cite{Yu2022} & Govern,Data & - & \like{10} & \like{10} & \like{10} \\
  \cline{2-7}
  & \cite{Boudagdigue2020} & Govern & \like{10} & \like{5} & - & - \\
  \cline{2-7}
  & \cite{10648633} & Govern,TrustNet & \like{10} & \like{5} & \like{5} & \like{5} \\
  \cline{2-7}
  & \cite{xiang2024blockchain} & Govern,TrustNet & \like{10} & \like{5} & \like{5} & \like{5} \\
  
  \hline
  \hline
  
  \multicolumn{7}{|c|}{\like{10}: Well supported; \like{5}: Partially supported; \like{0}: Not supported; -: Up to further discussion;} \\
  \multicolumn{7}{|c|}{Govern: Governance Layer; TrustNet: Trust Network Layer; Data: Data Layer} \\
  \hline
  \end{tabularx}
  \end{table*}

\section{Credential-based DTMS Evaluation}\label{evaluation}

Recent studies have proposed evaluation criteria for trust management. \cite{WangJie2022} lists factors such as subjectivity, dynamicity, context-awareness, privacy, scalability, robustness, overhead, explainability, and user acceptance. \cite{Ebrahimi2022} highlights decentralization, simplicity, adaptability, and efficiency for IoT, while \cite{Andersen2019} addresses delegation, verification, and offline management in decentralized authorization. Based on these, we propose four evaluation dimensions for credential-based DTMS: dynamicity, explainability, privacy, and scalability.


\subsection{Dynamicity} \label{DTMS dynamicity}

Trust in distributed systems is inherently dynamic, requiring mechanisms to manage its temporal evolution. This dynamicity arises from multiple factors:

\begin{itemize}
\item \textbf{Event-driven updates}: Trust relationships evolve continuously in response to contextual changes, emerging events, and interactions with other trust relationships. A robust DTMS must maintain trust data currency (including trust relationships, credentials, claims, and decisions) across the distributed network with minimal latency. This requires efficient event detection, propagation mechanisms, and update protocols.

\item \textbf{Transactional consistency}: Trust data updates frequently involve multiple distributed entities and credentials, demanding ACID-compliant transaction management \cite{Yu2009}. The system must guarantee atomicity to ensure all related updates succeed or fail together, consistency to maintain trust relationship validity, isolation to prevent interference between concurrent updates, and durability to persist trust state changes reliably.

\item \textbf{Network dynamics}: Modern distributed systems face continuous entity churn, with participants joining and departing frequently. DTMS must handle these topology changes gracefully, including special considerations for offline participants \cite{Andersen2019}. This encompasses efficient join/leave protocols, state synchronization mechanisms, and robust recovery procedures for reconnecting entities.

\item \textbf{Temporal consistency}: The system must maintain temporal consistency of trust relationships across distributed entities, implementing appropriate timestamp mechanisms, version control, and conflict resolution protocols to handle concurrent updates and ensure a coherent trust state across the network.
\end{itemize}

\subsection{Explainability}

A credential-based DTMS must provide comprehensive explainability across multiple dimensions:

\begin{itemize}
\item \textbf{Decision transparency}: The system should offer clear explanations for trust-driven decisions throughout the trust lifecycle, including relationship establishment, maintenance, and revocation. This encompasses both trust-specific decisions and application-level determinations such as access control, data acceptance, and entity classification. The explanation mechanism should provide both human-readable justifications and machine-processable logical derivations.

\item \textbf{Trust articulation}: Despite the inherent complexity of trust quantification, DTMS must provide mechanisms to explicitly capture and communicate trust reasoning. This includes maintaining evidence records, documenting decision rationales, and implementing formal trust policies that clearly define evaluation criteria and decision procedures. The system should transform implicit trust assumptions into explicit, verifiable rules.

\item \textbf{Verification and audit}: DTMS must support multi-level verification, enabling entities to validate both individual trust decisions and broader trust patterns. This requires maintaining verifiable evidence chains, implementing secure audit logging mechanisms, and supporting temporal analysis of trust evolution. Blockchain technology can provide immutable audit trails, while formal verification techniques can validate trust policy compliance.

\item \textbf{Reasoning framework}: The system should implement a structured reasoning framework that combines multiple trust factors, including direct experiences, recommendations, reputation scores, and contextual parameters. This framework must support both rule-based and probabilistic reasoning, with a clear articulation of how different factors are weighted and combined in trust computations.
\end{itemize}

\subsection{Privacy}

Privacy protection in credential-based DTMS requires a comprehensive security architecture:

\begin{itemize}
\item \textbf{Data minimization}: Following privacy-by-design principles \cite{privacybydesign}, the system must implement strict data minimization. This involves carefully defining essential trust data elements, implementing fine-grained disclosure controls, and supporting selective revelation of credentials. The system should provide mechanisms for entities to control their privacy exposure levels while maintaining necessary trust functionality.

\item \textbf{Privacy-preserving protocols}: Advanced cryptographic techniques, particularly ZKP \cite{Lauinger2021, WangJie2022}, should be employed to protect sensitive trust information. The system must support various privacy levels, from pseudonymous interactions to strongly authenticated exchanges, adapting to different use case requirements \cite{privacyspectrum}. Additional techniques such as homomorphic encryption and secure multi-party computation can enable trust computations over encrypted data.

\item \textbf{Attack resistance}: DTMS must incorporate robust defenses against privacy-related attacks, e.g., correlation attacks, inference attacks, and identity linking attempts \cite{10819652}. This requires careful protocol design, regular security analysis, and implementation of appropriate countermeasures such as mixing networks, dummy traffic, and privacy-preserving authentication mechanisms.

\item \textbf{Regulatory compliance}: The system should implement privacy controls aligned with relevant data protection regulations \cite{issa2022maritime}, supporting features such as data portability, right to erasure, and consent management. This includes maintaining privacy policies, implementing data lifecycle management, and providing transparency about data usage.
\end{itemize}

\subsection{Scalability}

Scalability in credential-based DTMS encompasses multiple critical aspects:

\begin{itemize}
\item \textbf{Performance scaling}: The system must maintain consistent performance under increasing loads, including growth in network size, event frequency, and relationship complexity \cite{WangJie2022}. This requires efficient data structures, optimized protocols, and careful resource management. The architecture should avoid centralized bottlenecks and support horizontal scaling of management functions.

\item \textbf{Algorithmic efficiency}: Core trust operations (e.g., establishment, verification, revocation) must employ efficient algorithms with acceptable computational complexity. This includes optimizing graph-based trust path discovery \cite{Franceschi2021}, implementing efficient game-theoretic strategies \cite{Boudagdigue2020}, and minimizing cryptographic overhead \cite{Lauinger2021}. Algorithm selection should consider both time and space complexity relative to network parameters.

\item \textbf{Resource optimization}: The system must efficiently manage computational, storage, and network resources. This includes implementing caching strategies, compression techniques, and adaptive resource allocation. Special attention should be paid to blockchain scalability challenges \cite{Salve2022}, implementing appropriate consensus mechanisms and data management strategies.

\item \textbf{Architectural scalability}: The system architecture must support seamless growth, including mechanisms for dynamic resource allocation, load balancing, and fault tolerance. This encompasses both vertical scaling (increasing individual node capacity) and horizontal scaling (adding more nodes), with appropriate consistency and synchronization protocols.
\end{itemize}

Our evaluation framework provides key advantages for assessing credential-based DTMS research through a clear three-level scale (well supported (\like{10}), partially supported (\like{5}), not supported (\like{0})) and four fundamental dimensions: dynamicity, explainability, privacy, and scalability. As demonstrated in Table \ref{evaluation_table}, we evaluate the papers from Section \ref{survey}, marking some aspects with "-" where further discussion is needed. This evaluation approach effectively reveals support levels across papers and identifies both achievements and areas for improvement in current research, offering valuable insights for future directions.

\section{Challenges and Future Directions}\label{challenge}

This section explores the critical challenges and future research directions in DTMS. Building upon our previous discussion of DTMS architectures, trust models, and implementation approaches, we identify key areas requiring further investigation and development to enhance the robustness, scalability, and practical applicability of these systems.

\subsection{Architectural Challenges}
The architectural design of DTMS presents fundamental challenges that must be addressed for successful implementation and deployment. These challenges span multiple layers, from governance and trust network management to data handling and storage.

\subsubsection{Governance Layer.}
The governance layer in DTMS must balance centralized control with entity autonomy~\cite{wang2025understanding,yu2023leveraging,tang2025decentralised}. Existing systems typically adopt either strict trust schemas or fully decentralized models. A hybrid approach could enforce baseline policies while supporting entity-specific customization, enabling runtime security validation, policy consistency checks, and flexible standard evaluation. Trust anchor implementations differ across domains, such as cluster heads in IoV, root certificate authorities in PKI, and trusted third parties in privacy systems. Each presents trade-offs in scalability, security, and overhead. Selection should consider network topology, trust propagation, and verification protocols to inform future DTMS designs.



\subsubsection{Trust Network Layer.}
RT management and backward chain discovery mechanisms must evolve for complex scenarios. Modern applications like supply chain management involve intricate certificate hierarchies and diverse relationships, creating multi-dimensional trust graphs with heterogeneous nodes and edges. Advanced graph modeling must address relationship representation, path discovery, and trust metrics. Optimization strategies for distributed environments remain crucial areas for investigation.

\subsubsection{Data Layer.}
Trust graphs present significant data management challenges requiring robust architecture for efficient operations and data integrity. ZKP protocols enable selective information disclosure for trust verification, while advanced indexing optimizes graph operations. Blockchain adoption as a data layer requires research in storage optimization, cost-effective management, and efficient query processing.

\subsection{Implementation Challenges}
The practical implementation of DTMS faces several technical challenges related to the adoption of emerging technologies such as blockchain and decentralized identifiers. These challenges require innovative solutions to ensure system efficiency, scalability, and security.

\smallskip
\subsubsection{Using Blockchain (transactional update)}
Blockchain technology, while providing immutable audit trails for trust evolution, presents significant challenges in managing dynamic trust relationships. Several architectural approaches address these challenges:

\begin{itemize}
    \item \textbf{State channels}: Facilitate efficient off-chain certificate updates while maintaining on-chain verification integrity, significantly reducing transaction costs and latency (e.g., MASS network implementation \cite{issa2022maritime});
    \item \textbf{Sharding}: Enhance system throughput through parallel processing of cross-domain authentication tasks, improving scalability in large-scale deployments (e.g., \cite{8954616} double-layer architectural model);
    \item \textbf{Pruning mechanisms}: Implement data retention policies to manage storage overhead while preserving essential trust history (e.g., certificate revocation);
    \item \textbf{Consensus optimization}: Develop specialized consensus protocols that balance security requirements with performance demands in trust management.
\end{itemize}

The integration of hybrid architectures, combining blockchain's immutability with traditional distributed databases' performance characteristics, offers promising solutions for optimizing system performance while maintaining trust guarantees.

\smallskip
\subsubsection{Using DID/VC}
Decentralized Identifiers (DID) and Verifiable Credentials (VC) introduce revolutionary approaches to identity and trust management:

\begin{itemize}
    \item \textbf{Cross-system interoperability}: DID's standardized URI format enables seamless entity identification across diverse domains, facilitating unified trust frameworks (cf. \cite{yu2023crosschainparentchainmultiple} IoT naming convention);
    \item \textbf{Selective disclosure}: Advanced ZKP-based VC implementations support granular privacy control in trust verification processes (cf. DART framework~\cite{Franceschi2021});
    \item \textbf{Lifecycle management}: Sophisticated mechanisms for managing dynamic relationships between DIDs and trust anchors, including anchor rotation and credential updates (e.g., BBRM's RCA election \cite{Sarker2021});
    \item \textbf{Resolution efficiency}: Optimization of DID solutions to minimize latency in large-scale networks.
\end{itemize}

Key challenges include optimizing DID document synchronization, developing efficient VC revocation for large-scale deployments, and establishing standardized interoperability across diverse DID systems.

\section{Conclusion}\label{conclusion}

This survey presents a comprehensive review of DTMS, covering fundamental concepts, key components, and emerging applications. We analyzed trust models, consensus mechanisms, and security frameworks that underpin DTMS, emphasizing their significance in modern digital ecosystems.

We demonstrated DTMS's vital role in facilitating trusted decentralized interactions, especially in blockchain and peer-to-peer systems. While identifying challenges like scalability and privacy, we explored potential solutions for advancing decentralized trust management. This work provides researchers and practitioners with a structured reference for developing secure decentralized systems.

\bibliographystyle{IEEEtran}
\bibliography{IEEEabrv}


\end{document}